\documentclass[12pt]{article}
\usepackage{amsmath}
\usepackage{graphicx}
\usepackage{enumerate}
\usepackage{natbib}
\usepackage{url} 

\usepackage{enumerate}
\usepackage{blkarray}
\usepackage{stackrel}
\usepackage{url}
\usepackage[top=1in, bottom=1in, left=1in, right=1in]{geometry} 

\usepackage{caption,subcaption,graphicx,psfrag,epsf,cases,empheq,color,amssymb,amsmath,graphicx,float,booktabs,amsthm,bm,bbm,multirow,threeparttable,textcomp,adjustbox,setspace,natbib,amsthm,soul,cancel,mathrsfs,euscript,amsfonts,xcolor,tikz,bm,blkarray,stackrel}
\usepackage{enumitem}
\usepackage{threeparttable}
\usepackage{times}
\usepackage{bm}
\usepackage{natbib}
\usepackage{siunitx}
\usepackage{amssymb}
\usepackage{blkarray}
\usepackage{stackrel}
\usepackage[plain,noend]{algorithm2e}
\usepackage{bm}
\usepackage{float}
\usepackage{tikz}
\usepackage{colortbl}
\usepackage{booktabs}
\usepackage{pbox}
\usepackage{longtable}
\usepackage{multirow}
\usepackage{amsfonts}
\usepackage{xr}
\usepackage{xcolor, soul}
\usepackage[hidelinks]{hyperref}
\theoremstyle{plain}

\newtheorem{theorem}{Theorem}

\newtheorem{assumption}{Assumption}

\usepackage{geometry}
\usepackage{pdflscape}
\usepackage[title]{appendix}
\usepackage{stackengine}

\newtheorem*{assumption*}{\assumptionnumber}

\providecommand{\assumptionnumber}{}

\newtheorem*{example*}{\examplenumber}
\providecommand{\examplenumber}{}


\def\T{{ \mathrm{\scriptscriptstyle T} }}

\providecommand{\assumptionnumber}{}

\def\bSig\mathbf{\Sigma}
\def\myinc{}

\newcommand{\bbeta}{\beta}
\newcommand{\bgamma}{\gamma}

\newcommand{\hatincthetak}{\tilde{\btheta}_{\myinc k}}

\newcommand{\hatincgammak}{\tilde{\gamma}_{\myinc k}}
\newcommand{\hatincgammaone}{\tilde{\gamma}_{\myinc 1}}
\newcommand{\hatincgammaj}{\tilde{\gamma}_{\myinc j}}

\newcommand{\hatincbetak}{\tilde{\bbeta}_{\myinc k}}
\newcommand{\hatincbetaone}{\tilde{\bbeta}_{\myinc 1}}

\newcommand{\hatincbetaj}{\tilde{\bbeta}_{\myinc j}}

\newcommand{\hatinczetak}{\tilde{\zeta}_{\myinc k}}
\newcommand{\hatinczetaone}{\tilde{\zeta}_{\myinc 1}}

\newcommand{\hatinczetaj}{\tilde{\zeta}_{\myinc j}}

\newcommand{\hatinczetakminus}{\tilde{\zeta}_{\myinc k-1}}

\newcommand{\hatinczetatwo}{\tilde{\zeta}_{\myinc 2}}

\newcommand{\oraclethetak}{\hat{\btheta}_{N_k}}
\newcommand{\oraclethetaK}{\hat{\btheta}_N} 
\newcommand{\oraclebetaK}{\hat{\bbeta}_{N}} 
\newcommand{\oraclebetak}{\hat{\bbeta}_{N_k}} 

\newcommand{\Hnjloc}{{H}_{j}}
\newcommand{\Vnjloc}{{V}_{j}}
\newcommand{\Hone}{{H}_{1}^{}}

\newcommand{\Vone}{{V}_{1}^{}}

\newcommand{\colsa}{\textsc{colsa}}
\def\CP{\textsc{CP}}

\def\ESE{\textsc{ESE}}
\def\ASE{\textsc{ASE}}
\def\MSE{\textsc{MSE}}

\def\AARB{\textsc{ARB}}

\def\T{{ \mathrm{\scriptscriptstyle T} }}
\newcommand{\ki}{{ki}}
\newcommand{\nk}{{n_k}}
\newcommand{\keone}{{k=1}}
 \newcommand{\lkpartialbeta}{{\ell_k^{\scriptstyle{partial}}(\bbeta)}}

\newcommand{\sievespace}{{{\Theta}_N^p}}
\newcommand{\sievespacenone}{{\Theta_{n_1}^p}}

\newcommand{\sievespacenk}{{\Theta_{N_k}^p}}

\newcommand{\btheta}{{{\theta}}}

\stackMath
\newcommand\tsup[2][2]{%
 \def\useanchorwidth{T}%
  \ifnum#1>1%
    \stackon[-.5pt]{\tsup[\numexpr#1-1\relax]{#2}}{\scriptscriptstyle\sim}%
  \else%
    \stackon[.5pt]{#2}{\scriptscriptstyle\sim}%
  \fi%
}
\date{}
\begin{document}
	\def\spacingset#1{\renewcommand{\baselinestretch}%
		{#1}\small\normalsize} \spacingset{1}
	\title{\Large\bf Privacy enhanced collaborative inference in the Cox proportional hazards model for distributed data}
	\author{Mengtong Hu, Xu Shi, and Peter X.-K. Song \\
		Department of Biostatistics, University of Michigan}
	\maketitle
	
	\bigskip
	\begin{abstract}
Data sharing barriers are paramount challenges arising from multicenter clinical studies where multiple data sources are stored in a distributed fashion at different local study sites. Particularly in the case of time-to-event analysis when global risk sets are needed for the Cox proportional hazards model, access to a centralized database is typically necessary. Merging such data sources into a common data storage for a centralized statistical analysis requires a data use agreement, which is often time-consuming.  Furthermore, the construction and distribution of risk sets to participating clinical centers for subsequent calculations may pose a risk of revealing individual-level information. 
We propose a new collaborative Cox model that eliminates the need for accessing the centralized database and constructing global risk sets but needs only the sharing of summary statistics with significantly smaller dimensions than risk sets. Thus, the proposed collaborative inference enjoys maximal protection of data privacy. We show theoretically and numerically that the new distributed proportional hazards model approach has little loss of statistical power when compared to the centralized method that requires merging the entire data. We present a renewable sieve method to establish large-sample properties for the proposed method. We illustrate its performance through simulation experiments and a real-world data example from patients with kidney transplantation in the Organ Procurement and Transplantation Network (OPTN) to understand the factors associated with the 5-year death-censored graft failure (DCGF) for patients who underwent kidney transplants in the US.
	\end{abstract}
	
	\vfill
	
	\newpage
	\spacingset{1.45} 
	
\section{Introduction}

\label{sec:intro}
Developing effective methods for integrating data from multiple clinical sites under a certain collective agreement of collaboration has gained growing interest among researchers, as it offers several significant advantages, including data enrichment for improving demographic diversity and expedition of scientific discoveries. 
Performing statistical analysis such as survival analysis in such settings becomes challenging when there are substantial barriers to directly sharing individual-level data across multiple clinical sites due to data privacy requirements, and the potentially time-consuming institutional Institutional Review Boards (IRBs) approval process. In this paper, we focus on addressing the challenges of developing distributed methods for modeling time-to-event data. Time-to-event outcome captures both the occurrence of the event and the time at which the event happens. This information is critical for understanding the progression of diseases in patients, particularly for events like deaths, disease relapses, or developments of adverse side effects.  The Cox proportional hazards model is one of the standard methods for understanding the association between patients' characteristics or potential treatments and time-to-event data~\citep{cox1972regression}.  

There has been substantial interest in developing distributed Cox regression models for advancing the understanding of diseases and expediting drug discoveries across multiple sources or health networks, without compromising the privacy of individual patients. WebDISCO was introduced by \cite{lu2015webdisco} for estimating the hazard ratios in the Cox proportional model using partial likelihood in a distributed fashion without sharing patient-level data. In WebDISCO, each local center communicates aggregated information and updated hazard ratio estimates with a central analysis server multiple times until a global maximum likelihood estimate for the hazard ratio is obtained.  The iterative nature of WebDISCO makes it time-consuming to implement in practice. Thus, many recent developments focus on the non-iterative implementation of the distributed partial likelihood which greatly shortens the execution time without losing noticeable accuracy or statistical efficiency~\citep{imakura2023dc,duanlocaltoglobal,li2022distributed,
lu2021multicenter}. A summary of relevant methods is listed in Table \ref{tab:methods} in Appendix 3. 

All the aforementioned distributed Cox methods focus on maximizing partial likelihood for parameter estimation, which require the sharing of event times across sites. This requirement stems from the need for partial likelihood-based methods to construct risk sets. These sets consist of at-risk patients who have not experienced the event of interest at a specific time $t$ and whose event or censoring time is greater than or equal to $t$. We denote the risk set at time $t$ as $R_t = \{i : Y_i \geq t\} $ where $Y_i$ is the observed event or censoring time for the $i$th individual. Suppose that there are $d$ distinct observed event times ranked across all sites, denoted by $t_{1} > t_{2}>t_{3}>\cdots >t_{d}$. Each site is provided with a copy of the ranked $d$ event times and determines which individuals from the local site belong to the risk set at each distinct time point. 
Once the risk sets have been constructed, aggregated summary statistics at the risk set level are shared among different sites to calculate the hazard ratios. Although sharing data at the risk-set level may seem preferable to sharing individual-level data, it's important to note that extracting individual-level covariate information can be straightforward using simple algebra. This is particularly true if the difference between two adjacent risk sets such as  $R_{t_m}$ and  $R_{t_{m-1}}$, contains only one individual from the local site.  This case is common in practice, especially when there are few or no ties in the event times. Thus, sharing event times and risk-set information presents challenges in privacy-preserving distributed learning~\citep{brink2022distributed,andreux2020federated}. 

To avoid sharing individual-level event times and constructing risk sets, meta-analysis is a frequently employed approach for estimating and drawing inferences about hazard ratios from site-specific summary statistics. A meta-estimation of hazard ratios can be calculated by inverse-variance weighted site-specific hazard ratios. However, limitations in real-world applications can compromise the statistical efficiency of classical meta-analysis. First, meta-analysis may suffer from data attrition due to varying recruitment capacity particularly when the event of interest does not happen frequently. Low event counts can lead to undefined or unreliable Cox model local estimates, leading to the exclusion of these sites from meta-analysis, thereby diminishing its statistical power~\citep{hu2024}. Of note, the convergence rate of the meta-estimation is at the sample size of the smallest data batch rather than the cumulative sample size~\citep{zhou2020sinica}. Furthermore, meta-analysis does not require that the nuisance parameters in the baseline hazard function be identical across sites and thus can lose statistical efficiency when there is no evidence suggesting nuisance parameters vary among different sites~\citep {lin2010relative}.

This paper focuses on developing a more reliable and inclusive privacy-preserving time-to-event model through effective management of data-sharing across a network of participating sites without constructing risk sets. We term this new approach as \emph{Collaborative Operation of Linked Survival Analysis} (\colsa).
By transforming the task of maximizing a partial likelihood function into maximizing a regular likelihood function, and by sharing aggregated summary statistics at the site level instead of the risk-set level, we can minimize the risk of disclosing individual information. 
The hazard ratios, along with their inferential statistics, are sequentially updated across sites using a renewable estimation method originally developed for solving generalized linear models in online learning settings~\citep{Luo_2020}. This incremental updating approach, differing from traditional meta-analytic methods that rely on parallelized operations, offers the advantage of not requiring asymptotically large sample sizes at each participating site.

To avoid using partial likelihood, we propose directly estimating the baseline hazard in proportional hazards models. \cite{wu2021online} approximates the cumulative baseline hazard function using a piecewise constant function in a streaming survival data setting, hereinafter referred to as the Online survival method. 
For an accurate approximation of the cumulative baseline hazard using the piecewise constant approach, a sufficiently fine partition of the event time is necessary. However, as the partition becomes finer, there is a substantial increase in the number of parameters that must be estimated. As a result, a much larger sample size is required to estimate all parameters accurately. Meanwhile, \cite{wu2021online} assumes the presence of a large sample size at each local site necessary to establish consistency in the estimated hazard ratios for each data block. 
We propose using a Bernstein polynomial function to estimate the logarithm of the baseline hazard function in the Cox regression model. Employing a cubic Bernstein basis polynomial offers a flexible approach to capturing the underlying hazard function, without imposing strong assumptions about its form. The complexity of a spline function is usually determined by the number of knots or the order of the polynomial, which sets the degrees of freedom allocated for modeling the hazard function. Extensive simulation studies have shown that the estimated hazard ratio values are largely insensitive to the degrees of freedom used for the specification of the baseline hazard when the knots are placed at equally distributed quantiles of the log of the event times~\citep{rutherford2015use}. The estimation and inference procedure of the polynomial-based Cox models is based on the sieve maximum likelihood estimation~\citep{shen1998propotional,shen1994convergence}. A large number of existing studies in the broader survival literature have adopted polynomial splines and sieve maximum likelihood to a variety of settings such as interval-censored data~\citep{zhou2017sieve,huang1997sieve,zhang2010spline}, panel count data ~\citep{wellner2007two,lu2009semiparametric}, and bundled parameters estimation ~\citep{ding2011sieve,zhao2017sieve,tang2023survival}. Following the same ideology, we incorporate sieve maximum likelihood estimation into the distributed Cox regression setting within  \colsa\ .

To elucidate key elements and insights of the proposed \colsa\ methodology, we investigate the biological and socioeconomic variables associated with death-censored graft failures among patients who had kidney transplants in the US.  Applying  \colsa\ methodology to the analysis is particularly meaningful when the goal is to draw conclusions on the entire population in a national-level analysis, where none of the regions or states has the capability to gather data on race, ethnicity, and health conditions for the entire US population of kidney transplant donors and recipients. Therefore, to draw conclusions about the entire US population without extensively sharing individual-level data, a decentralized approach for analyzing time-to-event data is desired. We obtain the relevant data from the Scientific Registry of Transplant Recipients (SRTR) collected by the Organ Procurement and Transplantation Network (OPTN). Within the OPTN network, there are 11 designated OPTN geographic regions to administer transplant operations across the US.  We consider the case where patient data from individual regions cannot be pooled or sent to a central data repository. 
The organization of this paper is as follows. Section~\ref{sec:Methodology} introduces the proposed \colsa\ framework. Section~\ref{proj2:sec:Asymptotics} establishes the theoretical guarantees for our proposed methods. We illustrate our proposed methods with simulation studies in Section~\ref{sec:simulation} and an application to the OPTN data example in Section~\ref{sec:data application}. We make some concluding remarks in Section~\ref{sec: discussion}.

\section{Methodology }\label{sec:Methodology}
\subsection{Problem set-up and partial likelihood }
Consider a random sample of $N$ individuals independently sampled from $K$ clinical sites, indexed by $k = 1, \cdots , K$, each site having a sample size of $n_k$. 
For the $i$th individual from site $k$, we observe a vector of baseline covariates $X_\ki$, and an observed time $Y_{\ki} := min(T_{\ki},C_{\ki})$, where $T_{\ki}$ and $C_{\ki}$ are the event time and censoring time, respectively. Let $\Delta_{\ki} := I(T_{\ki}\leq C_{\ki})$ be the event indicator. We denote the $k$th site-specific data as $\mathcal{O}_k=\{\mathcal{O}_{ki} = (X_{\ki},Y_{\ki},\Delta_{\ki}): i\in  \{1,\cdots,\nk\}\}$. We assume that 
 $\mathcal{O}_{ki}$ for $ i\in  \{1,\cdots,\nk\} $ and $ k \in  \{1,\cdots,K\}  $ are independent and identical observations under the same underlying population and the censoring is noninformative. The Cox proportional hazards model is defined through the following hazard function specification
\begin{align*}
  \lambda(t;X) = \lambda_0(t)\exp(X^\T\bbeta )
\end{align*}
where $\lambda(\cdot)$ is a hazard function of time $t$ and $\lambda_0(t)$ is the unspecitifed baseline hazard function and $\beta$ is the regression parameter vector. The overall logarithm of the partial likelihood function of the Cox model can be written as the sum of the logarithm of the partial likelihood from all $K$ sites, that is
\begin{align*}
 \ell^{\scriptstyle{partial}}(\bbeta) = \sum_{\keone}^{K} \lkpartialbeta = \sum_{\keone}^{K} \sum_{i=1}^{n_k}\Delta_{\ki}\ln
    \left\{ 
    \frac{\exp( X_{\ki}^\T\bbeta)}
    { \left[\sum_{k^\prime =1 }^K \sum_{i^{\prime}=1}^{n_{k^{\prime}}}I(Y_{k^\prime i^\prime}\geq Y_{\ki})\exp( X_{k^\prime i^{\prime}}^\T\bbeta)\right]}   
    \right\}\\
\end{align*}
where $\lkpartialbeta$ is the log partial likelihood of the $k$th site. The risk set, i.e., the individuals in the denominator, contains the set of all subjects who at risk at time $Y_{\ki}$ which can also be denoted by $R_{Y_{\ki}}$. We further denote the subset of subjects in $R_{Y_{\ki}}$ who are from the site $k^{\prime}$ as $R^{k^{\prime}}_{Y_{\ki}}$. We re-express the log partial likelihood of the $k$th site 
\begin{align}\label{e1}
\lkpartialbeta = 
\sum_{i=1}^{n_k}\Delta_{\ki}\ln
    \left\{ 
    \frac{\exp(X_{\ki}^\T\bbeta )}
    { \left[\sum_{k^\prime =1 }^K \sum_{i^\prime \in R^{k^\prime}_{Y_{\ki}}}\exp( X_{k^\prime i^{\prime}}^\T\bbeta)\right]}   
    \right\}
\end{align}

Since the indicator $\Delta_{\ki}$ in equation \eqref{e1} is non-zero only for subjects with observed event times, it is necessary to construct the risk sets at all these observed event times but not at censoring times. With a slight abuse of notation, denote a total of $d$ observed unique event times across all sites according to their relative order by $y_{(1)}  > \cdots >y_{(m)}> \cdots >y_{(d)}$. The number of unique event times $d$ is less than or equal to the total number of observed times, $N$, depending on the count of individuals experiencing events and the presence of tied event times. In distributed partial likelihood literature, constructing $\lkpartialbeta$ requires a distributive computation of the denominator across multiple sites. For example, at each site, the determination of membership in $R^{k^\prime}_{y_{(m)}}$ is made, followed by aggregating individual-level data from $R^{k^\prime}_{y_{(m)}}$ to calculate corresponding summary statistics ~\citep{imakura2023dc,duanlocaltoglobal,li2022distributed,
lu2021multicenter}. We argue that risk-set level data sharing is risky when two adjacent risk sets, say $R^{k^\prime}_{y_{(m)}}$ and $R^{k^\prime}_{y_{(m-1)}}$, being differed by only a few subjects
; adversarial attackers can leverage differences in summary statistics to reversely infer raw covariate information~\citep{brink2022distributed,andreux2020federated}.

\subsection{Using Bernstein basis polynomials to approximate log baseline hazard}
Let $ \int_{0}^t \lambda_0(u)du$ be the true cumulative baseline hazard. The log-likelihood function 
for individual $i$ from site $k$ based on the observed data is 
\begin{align}
    \ell(\bbeta, \lambda_0;\mathcal{O}_{\ki})& = \Delta_{\ki} \{X_{\ki}^\T \bbeta +\log \lambda_0(Y_{\ki})\}-\left\{\int_{0}^{Y_{\ki}}\lambda_0(s)ds \right\} \exp({X_{\ki}^\T \bbeta}).
    \label{eq:ll}
\end{align}
Let $g(t)$ be the logarithm of the baseline hazard function and $\btheta = (\bbeta,g)$, the individual log-likelihood function of $\btheta$ based on equation \eqref{eq:ll} is given by 
\begin{align*}
            \ell(\btheta; \mathcal{O}_{\ki})& = \Delta_{\ki} \{X_{\ki}^\T \bbeta +g(Y_{\ki})\}-\left\{\int_{0}^{Y_{\ki}} \exp\{g(s)\}ds\right\}\exp({X_{\ki}^\T \bbeta}).
\end{align*}
Let $a$ and $b$ be the lower and upper bounds of the observed times and let $\mathcal{G}$ denote the collection of bounded and continuous functions $g$ on $[a,b]$. Let $\mathcal{B} \subset \mathbb{R}^r$ be the space of feasible regression parameters at which the model is defined. We consider the estimation of $\btheta $ defined in the parameter space of $\Theta = \mathcal{B} \times \mathcal{G}$. Directly maximizing $\sum_{k=1}^{K} \sum_{i=1}^{n_k}  \ell(\btheta; \mathcal{O}_{\ki})$ in the space is not feasible as it involves infinite-dimensional parameters.   Following sieve maximum likelihood estimation, we propose to use polynomials to approximate $g$. Specifically, we choose Bernstein basis polynomials given that they do not require the specification of interior knots and that they have optimal shape-preserving properties and stability and stability among all approximation polynomials \citep{PENA19975,bernsteinFarouki}. On the interval $[a,b]$, the Bernstein polynomials of degree $p$ are defined as 
\begin{align*}
    B_j(t,p) =  \binom{p}{j} \left(\frac{t-a}{b-a}\right)^j \times \left(1-\frac{t-a}{b-a}\right)^{p-j}, j = 0, \cdots, p,
\end{align*}
with the degree $p = o(N^\nu)$ for some $\nu \in (0,1)$. 
Let $\mathcal{G}^p_N $ be the space of $p$-degree Bernstein polynomials and any polynomial function $s(t)$ in the space $\mathcal{G}^p_N$ 
can be written as a linear combination of Bernstein basis $ B_j(t,p) $ with $j = 0, \cdots, p$ such that 
$$s(t) = \sum_{j=0}^{p}\gamma_jB_j(t),\mbox{ }  \sum_{j=0}^{p}\vert \gamma_j\vert \leq M_N$$ with  $p$ omitted in the basis $B_j$ for simplicity and the size of the coefficients of the Berstein basis $\gamma = (\gamma_j)_{j=0\cdots p} $ is controlled by $M_N = O(N^\alpha)$ with $\alpha$ being a positive constant~\citep{lorentz2012bernstein,zhou2017sieve}. We define the sieve space as $\sievespace = \mathcal{B} \times  \mathcal{G}^p_{N}$ whose size is controlled by $M_N$. \textcolor{black}{According to the Weierstrass approximation theorem, every continuous function defined on a closed real-valued interval can be uniformly approximated by a polynomial function. In addition, under the suitable smoothness assumption (\ref{a1:continuity} in Appendix \ref{proj2:appx1}), the error of using some Bernstein Polynomial functions in $\mathcal{G}^p_N $ to approximate the true log baseline hazard function $g_0(\cdot) = \log \lambda_0(\cdot)$ is well controlled at the desirable rate shown in the comment in Appendix \ref{proj2:appx1}}. 
Therefore, we aim to find the sieve maximum likelihood estimator 
that maximizes the log-likelihood function $\sum_{k=1}^{K} \sum_{i=1}^{n_k}  \ell(\btheta; \mathcal{O}_{\ki})$ over  $\sievespace$. By employing the sieve maximum likelihood procedure, we turn the estimation problem about both finite-dimensional and infinite-dimensional parameters into a standard estimation that involves only finite-dimensional parameters $\bbeta$ and $\bgamma$ such that $(\hat{\bbeta}_N,\hat{\bgamma}_N)$ maximize the following sum of approximate log-likelihood:
\begin{align}
    \sum_{k=1}^{K} \sum_{i=1}^{n_k} \Delta_{\ki} \{\bbeta^TX_{\ki} +  \sum_{j=0}^{p}\gamma_jB_j(Y_{ki})\}-\left[\int_{0}^{Y_{\ki}} \exp\{ \sum_{j=0}^{p}\gamma_jB_j(s)\}ds \right ] \exp({X_{\ki}^\T \bbeta}).
\label{eq:approx}
\end{align}
Note that the constraint of $\sum_{j=0}^{p}\vert \gamma_j\vert \leq M_N$ is not necessary to be considered in the computation given that $M_N = O(N^\alpha)$  can be taken reasonably large for fixed $N$~\citep{zhou2020semiparametric}. We can obtain the score equations by taking the first-order derivatives of equation \eqref{eq:approx} with respect to $\bbeta$ and $\bgamma$ and setting them to zero, denoted by
\begin{align}
    \sum_{k=1}^{K} \sum_{i=1}^{n_k} U_{\ki}(\bbeta,\bgamma) = 0
        \label{eq:U}
\end{align}
such that $U_{ki}(\bbeta,\bgamma)$  
has the dimension of $r+p+1$.
The numerical implementation for finding the roots of the score equations can be done through the Newton-Raphson algorithm along with the Gaussian-quadrature method for approximating the integral part of the equation. 
The sieve maximum likelihood estimate $\hat{\btheta}_N$ described above is obtained from merging data from all $K$ sites in a centralized analysis and is thus regarded as the oracle estimator. The consistency of $\oraclethetaK$ and the asymptotic normality of $\oraclebetaK$ can be easily shown as discussed in Section \ref{proj2:sec:Asymptotics}. 

\subsection{Incremental hazard ratios estimator}

\label{proj2:subsec:incremental}
We consider a situation of practical importance where pooling data from multiple sites is prohibited. To address this data-sharing challenge, we propose an incremental estimation method, {Collaborative Operation of Linked Survival Analysis} (\colsa), which does not require sharing individual-level data but only certain summary statistics across sites. Note that \colsa\ is not derived in a parallel computing paradigm, commonly adopted in most existing solutions. We denote $\zeta = (\beta,\gamma) $ and given an order of study sites, $\zeta$ is sequentially updated over the first $k$ sites, beginning with $\hatinczetaone = (\hatincbetaone,\hatincgammaone)$ 
at study site 1 and ending with $\hatinczetak = (\hatincbetak,\hatincgammak)$ at the $k$th study site. We define $\hatincthetak$ as an incremental estimator obtained at site $k$ which is given by $\hatincthetak = (\hatincbetak, \sum_{j=0}^{p}\tilde{\gamma}_{\myinc k_{j}} B_j(t))$ where $\tilde{\gamma}_{\myinc k_{j}}$ is the $j+1$th element of $\tilde{\gamma}_{\myinc k}$ corresponding to  $ B_j(t)$. Obviously,  $\hatinczetaone =(\hatincbetaone,\hatincgammaone)$ 
is the same as a local estimator as a root of the estimating equation $\sum_{i=1}^{n_1} U_{1i}(\beta,\gamma)=0$ with the local data of $\mathcal{O}_1$, which is equivalent to maximize $\ell_1$ in the space of  $\sievespacenone = \mathcal{G}^p_{n_1} \times \mathcal{B}$.
We define $$\Hnjloc(\hatinczetaj)=\sum_{i = 1}^{n_j} {-\frac{\partial U_{ji}(\zeta)}{\partial \zeta^\T}\bigl\vert_{\zeta = (\hatincbetaj,\hatincgammaj)}},~ \Vnjloc(\hatinczetaj)=\sum_{i = 1}^{n_j} {U_{ji}(\zeta)U_{ji}^\T (\zeta)\bigl\vert_{\zeta=(\hatincbetaj,\hatincgammaj)}}$$ as the sensitivity and variability matrix, respectively, evaluated at a local site $j\in\{1,\dots,K\}$. 
The initial estimates of the sensitivity matrix and variability matrix 
$\{\Hone(\hatinczetaone)$, $\Vone(\hatinczetaone)\}$ with the local data $\mathcal{O}_1$, are also updated by \colsa. Notice that we assume $O(n_1) = O(N) $, ensuring the sieve space $\sievespacenone $ has the same order of size as $\sievespacenk $. 
After obtaining $\{\hatinczetaone$, $\Hone(\hatinczetaone)$, $\Vone(\hatinczetaone)\}$ from site 1,  \colsa\ passes these summary statistics to site $2$. 

We expand $\hatinczetaone$  to site 2 where $\hatinczetaone$ is updated to $\hatinczetatwo$ by solving the following estimating equation~\citep{Luo_2020}:
\begin{equation*}
\sum_{i = 1}^{n_2}U_{2i}(\hatinczetatwo)+
\Hone(\hatinczetaone)(\hatinczetaone-\hatinczetatwo)
=0,
\end{equation*}
Or equivalently, maximizing the following likelihood function
\begin{equation}\label{proj2:one}
    \sum_{i = 1}^{n_2}\ell(\hatinczetatwo;\mathcal{O}_{2i})+\frac{1}{2}
(\hatinczetaone-\hatinczetatwo)^\T H_1(\hatinczetaone)(\hatinczetaone-\hatinczetatwo)
,
\end{equation}
Repeating this sequential updating, \colsa\ can be carried out over a sequence of all sites to produce estimators and inferential quantities. 
In particular,  $\hatinczetakminus$ is updated to $\hatinczetak$ at site $k$ and $\hatinczetak$ satisfies the following estimating equation:  
\begin{equation}
\label{proj2:two}
\sum_{i = 1}^{n_k}U_{ki}(\hatinczetak)+\sum_{j=1}^{k-1}\Hnjloc(\hatinczetaj) (\hatinczetakminus-\hatinczetak)=0.
\end{equation}
 The estimating function in \eqref{proj2:two} 
consists of two parts: the first term $\sum_{i = 1}^{n_k}U_{ki}(\hatinczetak)$ is based on the local data at site $k$, and the second term assembles the cumulative summary statistics preceding from all previous $k-1$ sites. The Newton-Raphson algorithm is applied to numerically find the solution $\hatinczetak$. 

For statistical inference, we sequentially compute the cumulative sensitivity and variability matrices over the first $k$ sites evaluated at a given point estimate. For example, if we update the sensitivity and variability matrices along with the incremental estimates, then at site $k$, the sensitivity matrix is given by $ \sum_{j=1}^k\{\Hnjloc(\hatinczetaj)\} $ and the variability matrix is given by $ \sum_{j=1}^k\{\Vnjloc(\hatinczetaj)\}.$ Then we compute the estimated covariance matrix using the inverse of the variability matrix.

Notice that the degree of the polynomial $p$ is related to the total sample size $N$, and in practice, one may try a few candidates for $p$ and select the best model using the Akaike Information Criterion (AIC). However, previous simulation results show that the estimation results are robust to the choice of degree for Bernstein polynomials \citep{zhou2017sieve}. Similar results are also found in the simulation section below.

\subsection{Incremental survival curves estimation}\label{proj2:sec:method:surv}
Compared to the semi-parametric Cox model where the baseline hazard is not specified, the cumulative baseline hazard function can be directly estimated by \colsa\ using $\hatincgammak$ as $\sum_{j=0}^{p}\tilde{\gamma}_{\myinc k_{j}} B_j(t)$, denoted by $\Tilde{\Lambda}_k(t)$. The baseline survival function can then be estimated as $\exp[-\Tilde{\Lambda}_k(t)]$. Under the proportional hazards assumption, survival function estimates $\{\exp[-\Tilde{\Lambda}_k(t)]\}^{\exp({x^{\T}\hatincbetak})}$ for a specified covariate vector $x$  can be obtained. Unlike the construction of the Breslow or Nelson-Aalen estimate used in WebDISCO and related works, our proposed approach for estimating the survival function once again does not require the construction of risk sets. 

\section{Large-sample Properties}\label{proj2:sec:Asymptotics}
We discuss the large sample properties of the proposed incremental estimator $\hatincthetak$ as $N_k=\sum_{j=1}^k n_j \to \infty$ instead of $\emph{min}_{j \in \{1,\cdots ,K\}}n_j \to \infty$ in the parallel computing paradigm. This condition is satisfied when $n_j \to \infty$ at one of the sites, or when the number of sites $k \to \infty$ and the former is the focus of this paper. Although we describe the estimating procedure mostly in terms of $\beta$ and $\gamma$ in \ref{proj2:subsec:incremental}, $\hatincthetak$ is directly given by  $(\hatincbetak, \sum_{j=0}^{p}\tilde{\gamma}_{\myinc k_{j}} B_j(t))$. Under the proportional hazards model,  
the true value for $\hatincthetak$ is given by $\theta_0 = (\beta_0,g_0)$.
For any $\btheta_1 = (\bbeta_1,g_1)$ and  $\btheta_2 = (\bbeta_2,g_2)$ in the space of $\Theta$, define an $L_2$-metric:
$ \Vert\btheta_1-\btheta_2\Vert = \{\Vert\bbeta_1-\bbeta_2\Vert^2 + \Vert g_1-g_2\Vert_{\mathcal{G}}^2\}^{1/2}$ where $\Vert \cdot \Vert_{\mathcal{G}}$ denotes some norms in the space ${\mathcal{G}}$. For $\rho > 0$,
let $N_{\rho}(\btheta_0)=\{\btheta:\Vert\btheta-\btheta_0\Vert \leq \rho\}$,  be a compact neighborhood of size $\rho$ around the true value $\btheta_0$. We assume the following regularity conditions to guarantee the asymptotic results. 
\begin{assumption}[Regularity Conditions]\label{proj2:a:reg} Let $v$ be the first-order directional derivative at the direction $v \in V$ which is defined in Appendix 2, we assume the following conditions on $\Dot{\ell}(\btheta,\mathcal{O} ) [v]$  
\begin{enumerate}[label = (\alph*), ref = \ref{proj2:a:reg}(\alph*)]
\item \label{proj2:c:one}
The true value $\btheta_0$ is the unique solution to $E\{\Dot{\ell}(\btheta,\mathcal{O} ) [v]\}=0$.  
\item \label{proj2:c:two}
The directional derivative $\Dot{\ell}(\btheta,\mathcal{O} ) [v]$  is continuously differentiable for all $\btheta$ in the neighborhood $N_{\rho}(\btheta_0)$.
\item \label{proj2:c:four}
The gradient of the second-order directional derivative $\Ddot{\ell}(\btheta;\mathcal{O})[v,\bar{v}]$ for $v,\bar{v} \in V$ defined in Appendix 2 is continuous for all $\btheta \in$  $N_{\rho}(\btheta_0)$ and its smallest eigenvalue goes to infinity.
\end{enumerate}

\end{assumption}
Furthermore, the asymptotic properties of the proposed estimator $\hatincthetak$ 
will be established on the consistency of $\oraclethetak$ and the asymptotic normality of $\oraclebetak$. To guarantee the well-known properties of the oracle estimator, we require additional regularity conditions \ref{s1:a}-  \ref{a1:e} which can be found in Assumption 
\ref{sieve} in Appendix 1. 
By following the steps in \cite{zhou2017sieve} and \cite{chen2012interval}, the proofs of the asymptotics for the oracle estimator in the simple right-censoring scenario can be constructed very similarly. With regularity conditions ~\ref{s1:a} $-$~\ref{a1:e} and ~\ref{proj2:c:one} $-$~\ref{proj2:c:four}, we conclude the following asymptotic properties of our proposed estimator and the proofs can be found in Appendix 2.

\setcounter{equation}{0}
\begin{theorem}
\label{proj2:theorem1}
Under the regularity conditions~\ref{sieve} and ~\ref{proj2:c:one} $-$~\ref{proj2:c:four}, the \colsa\ estimator $\hatincthetak$ is consistent for the true value $\btheta_0$, i.e. $\hatincthetak\xrightarrow[]{}\btheta_0$ in probability, as $N_k \to \infty$. 
\end{theorem}

\begin{theorem}\label{proj2:theorem2}
Let  $\ell^{*}_{\bbeta_0}(\btheta_0,\mathcal{O}) $ defined in Appendix 2  be the efficient score function for $\bbeta_0$. 
 Supposed that Assumption \ref{sieve} hold and $\nu >(2q)^{-1}$ where $q$ is defined in \ref{a1:continuity}, then 
$N_k^{1/2}(\hatincbetak-\bbeta_0)$ converges in distribution to a mean zero normal random vector with variance-covariance matrix $I^{-1}(\bbeta_0)$ equal to the semiparametric efficiency bound of $\bbeta_0$, that is, 
$$\sqrt{N}_k(\hatincbetak-\bbeta_0)  
\xrightarrow{} \mathcal{N}(0,I^{-1}(\bbeta_0)), $$
where $I(\bbeta_0) = E\left[\ell^{*}_{\bbeta_0}(\btheta_0;\mathcal{O})\{\ell^{*}_{\bbeta_0}(\btheta_0;\mathcal{O})\}^{\T}\right]. $
\end{theorem}

\section{Simulation Experiments}\label{sec:simulation}
We evaluate the finite-sample performance of the proposed method with varying degrees of polynomials ($p = 2,3,\mbox{and } 4$), comparing it with the oracle estimation (i.e., the gold standard obtained by the centralized analysis), the classical meta-analysis
(the inverse-variance weighted meta method \citep{cochran1954combination}), and the Online survival method~\citep{wu2021online}. We first generate the full data under an assumed proportional hazards model and then split the data into $K$ subsets, one for a study site. We assume that the baseline hazard follows the two-component mixture Weibull distribution with a mixture probability of $0.5$, a scale parameter vector of $(10, 20)$, and a shape parameter vector of $(3, 5)$. 
To mimic the covariate distribution in the data example, we generate covariates $X_1$ to $X_4$ with some correlations: two continuous variables $X_1$ and $ X_2$ are drawn from a multivariate normal distribution with a mean vector of $(5,5)$ and a variance-covariance matrix equal to $(10,3;3,2)$ and one binary covariate $X_3$ is drawn from a Bernoulli distribution with a success probability of $0.8$. Motivated by the possible distribution of race and ethnicity variables commonly included in survival analysis (see data example for more details), $X_4$ is a four-level categorical variable with a multinomial probability vector of $(0.2, 0.2, 0.3, 0.3)$ or $(0.1, 0.2, 0.4, 0.5)$ for $X_3 =0$ and  $X_3 =1$ respectively. For convenience, we use $X$ to denote the vector of all five covariates. The true value of the log hazard ratio is $(0.15,-0.15,0.3,0.3,0.3,0.3)$. The time-to-event outcome $Y$ is generated through the cumulative hazard inversion method (\cite{simsurv,bender2005generating}). The censoring time $C$ is independently generated from an exponential distribution with a rate parameter of $6$ such that the event rate is controlled to be around $12\%$. To select the most optimal $p$, we use AIC based on the data from the first site. Simulation results are averaged over $500$ replications limited to those where all algorithms converge successfully in order to make a fair comparison. 

Table 1 represents the simulation results with data generated from the assumed Cox proportional hazards model above with the number of sites $K$ varying to be $6$, $20$, and $50$. The first six sites have fixed sample sizes of $1500$, $1500$, $1500$, $500$, $500$, and $500$, and the rest of the $K-6$ sites have equal sample sizes of $100$ which mimics the common scenario in practice where some participating sites have small sample sizes. The performance of \colsa\ across various polynomial degrees and the AIC-based \colsa\ is similar, indicating insensitivity to the degrees of the polynomial. This allows us to focus our discussion on comparing the various \colsa\ variations with other methods. Two bias metrics are used in assessing the accuracy of log hazard ratios estimation of different methods. $\AARB$ calculates the percentage of relative bias of log hazard ratios estimated by other methods with respect to the gold standard oracle estimates and $\MSE$ calculates the mean squared bias of estimates with respect to the true estimate $\bbeta_0$. Compared to $\colsa$, the meta estimate has at least twice the $\AARB$ for all scenarios, with the greatest $\AARB$ being $65.8\%$, which is more than $13$ times the $\AARB$ of $\colsa$ ($2.9 - 5.1\%$) for the categorical variable parameter $\bbeta_6$ when $K = 50$. \ASE\ is the average standard error and \ESE\ represents the empirical standard error for the sample and both metrics demonstrate the efficiency of each method. The differences in \ASE\ between \colsa\ and oracle are ignorable, while meta has consistently larger \ASE. Similar patterns are observed on \ESE. As the sample size grows with increasing $K$, the biases of oracle and \colsa~ decrease, and the estimates are approaching the true values with \MSE\ being as small as $1.43\times 10^{-4}$. Although \MSE\ also decreases for meta, it decreases at a much slower rate which is illustrated by the increasing $\AARB$. This is because some local sites fail to converge and do not contribute to the overall statistical power. The Online method with adaptive baseline hazards partition and bias correction has inferior performance compared to all other methods, as it is designed for large sample sizes and might not be suitable for the simulation setting we considered.
\begin{table}[]\footnotesize
\centering
\caption{
	Simulation results for $\bbeta_1$ and $\bbeta_6$ for varying numbers of sites.}\label{proj2:tab:simulation}
 \scalebox{0.76}{
\begin{tabular}{lllllllllllllllll}
\multicolumn{1}{c}{}                                             &       & \multicolumn{15}{c}{$n_1=1500,n_2 = 1500, n_3 = 1500, n_4 = 500, n_5 = 500, n_6 = 500$, $n_k = 100$ for $ k>6$, $P(X_3=1) = 0.8$}                                                          \\&       & \multicolumn{5}{c}{K = 6}  & \multicolumn{5}{c}{K = 20} & \multicolumn{5}{c}{K = 50} \\&       & \multicolumn{15}{c}{$\beta_1$}      \\
\multicolumn{1}{c}{\textit{Methods}}                             & $p$ & \multicolumn{1}{c}{\textit{\begin{tabular}[c]{@{}c@{}}\AARB\\ (\%)\end{tabular}}} & \multicolumn{1}{c}{\textit{\begin{tabular}[c]{@{}c@{}}\CP\\ (\%)\end{tabular}}} & \multicolumn{1}{c}{\textit{\begin{tabular}[c]{@{}c@{}}\MSE \\ $10^{-4}$\end{tabular}}} & \multicolumn{1}{c}{\textit{\begin{tabular}[c]{@{}c@{}}\ASE \\ $10^{-2}$\end{tabular}}} & \multicolumn{1}{c}{\textit{\begin{tabular}[c]{@{}c@{}}\ESE\\ $10^{-2}$\end{tabular}}} & \multicolumn{1}{c}{\textit{\begin{tabular}[c]{@{}c@{}}\AARB\\ $(\%)$\end{tabular}}} & \multicolumn{1}{c}{\textit{\begin{tabular}[c]{@{}c@{}}\CP\\ $(\%)$\end{tabular}}} & \multicolumn{1}{c}{\textit{\begin{tabular}[c]{@{}c@{}}\MSE \\ $10^{-4}$\end{tabular}}} & \multicolumn{1}{c}{\textit{\begin{tabular}[c]{@{}c@{}}\ASE \\ $10^{-2}$\end{tabular}}} & \multicolumn{1}{c}{\textit{\begin{tabular}[c]{@{}c@{}}\ESE \\ $10^{-2}$\end{tabular}}} & \multicolumn{1}{c}{\textit{\begin{tabular}[c]{@{}c@{}}\AARB\\ $(\%)$\end{tabular}}} & \multicolumn{1}{c}{\textit{\begin{tabular}[c]{@{}c@{}}\CP\\ $(\%)$\end{tabular}}} & \multicolumn{1}{c}{\textit{\begin{tabular}[c]{@{}c@{}}\MSE \\ $10^{-4}$\end{tabular}}} & \multicolumn{1}{c}{\textit{\begin{tabular}[c]{@{}c@{}}\ASE \\ $10^{-2}$\end{tabular}}} & \multicolumn{1}{c}{\textit{\begin{tabular}[c]{@{}c@{}}\ESE \\ $10^{-2}$\end{tabular}}} \\
\cellcolor[HTML]{FFFFFF}{\color[HTML]{222222} \textit{Oracle}}   & -     & 0.0                                                                                 & 95.6                                                                              & 2.63                                                                                   & 1.62                                                                                   & 1.62                                                                                  & 0.0                                                                                 & 96.1                                                                              & 1.99                                                                                   & 1.45                                                                                   & 1.41                                                                                   & 0.0                                                                                 & 95.8                                                                              & 1.44                                                                                   & 1.22                                                                                   & 1.19                                                                                   \\
\cellcolor[HTML]{FFFFFF}{\color[HTML]{222222} \textit{$\colsa$}} & 2     & 1.0                                                                                 & 95.8                                                                              & 2.66                                                                                   & 1.61                                                                                   & 1.63                                                                                  & 1.0                                                                                 & 96.4                                                                              & 2.01                                                                                   & 1.45                                                                                   & 1.42                                                                                   & 0.8                                                                                 & 96.3                                                                              & 1.47                                                                                   & 1.22                                                                                   & 1.19                                                                                   \\
$\colsa$                                                         & 3     & 0.7                                                                                 & 96.0                                                                              & 2.62                                                                                   & 1.61                                                                                   & 1.62                                                                                  & 0.6                                                                                 & 96.4                                                                              & 1.96                                                                                   & 1.45                                                                                   & 1.40                                                                                   & 0.5                                                                                 & 96.0                                                                              & 1.43                                                                                   & 1.22                                                                                   & 1.19                                                                                   \\
\cellcolor[HTML]{FFFFFF}{\color[HTML]{222222} \textit{$\colsa$}} & 4     & 0.5                                                                                 & 95.6                                                                              & 2.63                                                                                   & 1.61                                                                                   & 1.62                                                                                  & 0.5                                                                                 & 96.4                                                                              & 1.99                                                                                   & 1.45                                                                                   & 1.41                                                                                   & 0.4                                                                                 & 95.3                                                                              & 1.44                                                                                   & 1.22                                                                                   & 1.19                                                                                   \\
$\colsa$AIC                                                      & -     & 0.7                                                                                 & 95.8                                                                              & 2.64                                                                                   & 1.61                                                                                   & 1.63                                                                                  & 0.6                                                                                 & 96.1                                                                              & 1.96                                                                                   & 1.45                                                                                   & 1.40                                                                                   & 0.4                                                                                 & 95.5                                                                              & 1.44                                                                                   & 1.22                                                                                   & 1.19                                                                                   \\
\textit{Meta}                                                    & -     & 2.1                                                                                 & 95.2                                                                              & 2.96                                                                                   & 1.66                                                                                   & 1.71                                                                                  & 3.6                                                                                 & 95.4                                                                              & 2.62                                                                                   & 1.60                                                                                   & 1.59                                                                                   & 5.2                                                                                 & 95.3                                                                              & 2.41                                                                                   & 1.50                                                                                   & 1.52                                                                                  \\
Online                                                           & -     & 6.5                                                                                 & 90.7                                                                              & 3.42                                                                                   & 1.65                                                                                   & 1.62                                                                                  & 7.2                                                                                 & 92.7                                                                              & 3.38                                                                                   & 1.63                                                                                   & 1.59                                                                                   & 19.8                                                                                & 88.9                                                                              & 429.56                                                                                 & 1.63                                                                                   & 20.72                                                                                  \\&       &                                                                                     &                                                                                   &                                                                                        &                                                                                        &                                                                                       &                                                                                     &                                                                                   &                                                                                        &                                                                                        &                                                                                        &                                                                                     &                                                                                   &                                                                                        &                                                                                        &                                                                                        \\&       & \multicolumn{15}{c}{$\beta_6$}  \\
\multicolumn{1}{c}{\textit{Methods}}                             & $p$ & \multicolumn{1}{c}{\textit{\begin{tabular}[c]{@{}c@{}}\AARB\\ $(\%)$\end{tabular}}} & \multicolumn{1}{c}{\textit{\begin{tabular}[c]{@{}c@{}}\CP\\ $(\%)$\end{tabular}}} & \multicolumn{1}{c}{\textit{\begin{tabular}[c]{@{}c@{}}\MSE \\ $10^{-4}$\end{tabular}}} & \multicolumn{1}{c}{\textit{\begin{tabular}[c]{@{}c@{}}\ASE \\ $10^{-2}$\end{tabular}}} & \multicolumn{1}{c}{\textit{\begin{tabular}[c]{@{}c@{}}\ESE\\ $10^{-2}$\end{tabular}}} & \multicolumn{1}{c}{\textit{\begin{tabular}[c]{@{}c@{}}\AARB\\ $(\%)$\end{tabular}}} & \multicolumn{1}{c}{\textit{\begin{tabular}[c]{@{}c@{}}\CP\\ $(\%)$\end{tabular}}} & \multicolumn{1}{c}{\textit{\begin{tabular}[c]{@{}c@{}}\MSE \\ $10^{-4}$\end{tabular}}} & \multicolumn{1}{c}{\textit{\begin{tabular}[c]{@{}c@{}}\ASE \\ $10^{-2}$\end{tabular}}} & \multicolumn{1}{c}{\textit{\begin{tabular}[c]{@{}c@{}}\ESE \\ $10^{-2}$\end{tabular}}} & \multicolumn{1}{c}{\textit{\begin{tabular}[c]{@{}c@{}}\AARB\\ $(\%)$\end{tabular}}} & \multicolumn{1}{c}{\textit{\begin{tabular}[c]{@{}c@{}}\CP\\ $(\%)$\end{tabular}}} & \multicolumn{1}{c}{\textit{\begin{tabular}[c]{@{}c@{}}\MSE \\ $10^{-4}$\end{tabular}}} & \multicolumn{1}{c}{\textit{\begin{tabular}[c]{@{}c@{}}\ASE \\ $10^{-2}$\end{tabular}}} & \multicolumn{1}{c}{\textit{\begin{tabular}[c]{@{}c@{}}\ESE \\ $10^{-2}$\end{tabular}}} \\
\cellcolor[HTML]{FFFFFF}{\color[HTML]{222222} \textit{Oracle}}   & -     & 0.0                                                                                 & 95.0                                                                              & 189                                                                                    & 13.8                                                                                   & 13.7                                                                                  & 0.0                                                                                 & 95.6                                                                              & 152                                                                                    & 12.5                                                                                   & 12.3                                                                                   & 0.0                                                                                 & 95.3                                                                              & 119                                                                                    & 10.5                                                                                   & 10.8                                                                                   \\
\cellcolor[HTML]{FFFFFF}{\color[HTML]{222222} \textit{$\colsa$}} & 2     & 9.0                                                                                 & 95.6                                                                              & 185                                                                                    & 13.7                                                                                   & 13.6                                                                                  & 6.8                                                                                 & 95.9                                                                              & 148                                                                                    & 12.4                                                                                   & 12.2                                                                                   & 5.1                                                                                 & 94.8                                                                              & 114                                                                                    & 10.5                                                                                   & 10.7                                                                                   \\
\textit{$\colsa$}                                                & 3     & 7.6                                                                                 & 94.8                                                                              & 191                                                                                    & 13.8                                                                                   & 13.8                                                                                  & 6.1                                                                                 & 95.9                                                                              & 153                                                                                    & 12.4                                                                                   & 12.4                                                                                   & 4.2                                                                                 & 94.6                                                                              & 118                                                                                    & 10.5                                                                                   & 10.8                                                                                   \\
\cellcolor[HTML]{FFFFFF}{\color[HTML]{222222} \textit{$\colsa$}} & 4     & 6.3                                                                                 & 94.8                                                                              & 191                                                                                    & 13.8                                                                                   & 13.8                                                                                  & 4.3                                                                                 & 95.6                                                                              & 153                                                                                    & 12.4                                                                                   & 12.4                                                                                   & 2.9                                                                                 & 94.6                                                                              & 118                                                                                    & 10.5                                                                                   & 10.8                                                                                   \\
\textit{$\colsa$AIC}                                             & -     & 7.3                                                                                 & 94.8                                                                              & 191                                                                                    & 13.8                                                                                   & 13.8                                                                                  & 5.4                                                                                 & 95.6                                                                              & 153                                                                                    & 12.4                                                                                   & 12.4                                                                                   & 4.0                                                                                 & 94.6                                                                              & 118                                                                                    & 10.5                                                                                   & 10.8                                                                                   \\
\textit{Meta}                                                    & -     & 18.7                                                                                & 94.2                                                                              & 206                                                                                    & 14.2                                                                                   & 14.2                                                                                  & 34.4                                                                                & 92.7                                                                              & 217                                                                                    & 13.6                                                                                   & 13.3                                                                                   & 65.8                                                                                & 84.2                                                                              & 290                                                                                    & 12.5                                                                                   & 12.0                                                                                   \\
Online                                                           & -     & 59.3                                                                                & 81.0                                                                              & 402                                                                                    & 13.0                                                                                   & 16.6                                                                                  & 56.0                                                                                & 78.6                                                                              & 454                                                                                    & 12.9                                                                                   & 17.4                                                                                   & 78.5                                                                                & 76.5                                                                              & 4410                                                                                   & 12.8                                                                                   & 65.4                                                                                  
\end{tabular}}\begin{tablenotes}{\item $\AARB$\, Average absolute relative bias with respect to the oracle estimate; \CP, coverage probability; \MSE, mean sqaured biase; \ASE, mean estimated standard error of the estimates; \ESE\, empirical standard error. 
}\end{tablenotes}\end{table}
 \section{Data Application Example}\label{sec:data application}
The SRTR database includes patients who underwent kidney transplants in the US under OPTN. The goal of the analysis is to understand the factors associated with the 5-year death-censored graft failure (DCGF) for patients who underwent kidney transplants in the US. We restricted the analysis to a sample size of 62,684 patients who had first-time kidney transplants from 2003 to 2006 from all 11 geographic OPTN regions across the US. We further excluded 695 patients who died on the same day of the transplant likely due to surgical difficulties.

The overall 5-year DCGF rate is $13.5\%$ across all regions, with Region 6 having the lowest rate of $10.0\%$ for 1752 transplant cases and Region 2 having the highest rate of $16.4\%$ for 6200 transplant cases. To assess the performance of the distributed Cox model, we consider the case when recipients' protected health information such as demographics and recovery trajectories after transplant can not be transmitted across OPTN regions nor sent to a central analysis server. We adjust for the following covariates: age, gender, BMI, race, and ethnicity of donors and recipients, whether the living donor or deceased donor, diabetes and hypertension status of the recipients, and the human leukocyte antigen (HLA) matching type between donors and recipients. The proportion of missingness in the covariates is mild, thus we conduct a complete-case analysis including 48,766 participants with complete information.

We conduct \colsa\ without requiring subject-level data sharing. This analysis is particularly meaningful when the goal is to draw conclusions on the entire population in a national-level analysis, where none of the regions or states has the capability to gather data on race, ethnicity, and health condition for the entire US population of kidney transplant donors and recipients. Thus, a decentralized way to analyze time-to-event data is desired in order to draw conclusions about the entire US population.  
We also perform a centralized analysis of the pooled data as the gold standard for benchmarking, the classical meta-analysis based on state-specific estimates, and the Online survival updating procedure for comparison~\citep{wu2021online}. 

One of the data analysis challenges is the unequal proportions of different race and ethnicity distribution across states. The donors and recipients were grouped into four categories based on their race and ethnicity: White, Black, Hispanic, and other races. Figure 2 shows the distribution of race and ethnicity by region.   
\begin{figure}[h]
\includegraphics[width=16cm]{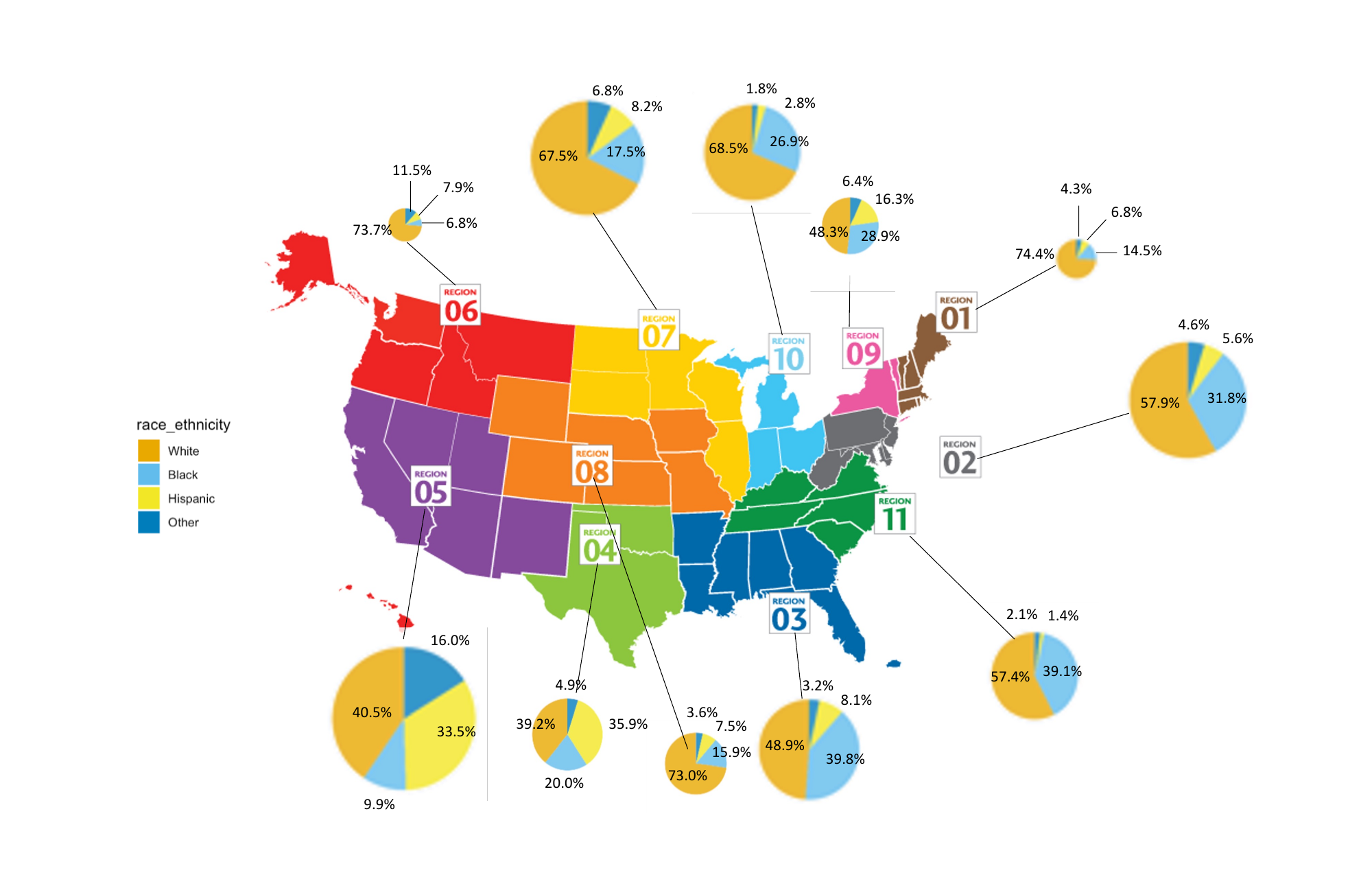}\label{fig:rarecp}
	\caption{A illustration of the distribution of race and ethnicity in different OPTN regions. The map with 11 regions is taken from OPTN webpage: \newline
 \textit{https://optn.transplant.hrsa.gov/about/regions}.}

\end{figure}

The little presence (less than 3\%) of individuals of specific race and ethnicity categories prevents us from getting reliable site-specific hazard ratios of recipients of those race groups in those regions.
Another challenge in the data analysis is that the ratios of sample sizes and numbers of covariates are low and the sample sizes of some smaller states are inadequate to provide numerically stable site-specific hazard ratio estimates. This data attrition is undesirable in classical meta-analysis.  
\begin{table}[]
\centering
\caption{{Hazard ratio estimates for DCGF from different methods.} }{
\resizebox{0.9\columnwidth}{!}{
\begin{tabular}{@{}lcccccccc@{}}
     & \multicolumn{2}{c}{Oracle}                  & \multicolumn{2}{c}{\colsa}         & \multicolumn{2}{c}{Meta}                              & \multicolumn{2}{c}{Online}                                \\ 
                                                                                        & HR                   & z-score               & HR                   & z-score                     & HR                       & z-score                     & HR                       & z-score                      \\
Recipient age                                                                           & 0.98                 & -21.30                & 0.98                 & \multicolumn{1}{l}{-21.21}
& \multicolumn{1}{l}{0.98} & \multicolumn{1}{l}{-21.42} & \multicolumn{1}{l}{0.98} & \multicolumn{1}{l}{-19.59}   \\
Donor age                                                                               & 1.02                 & 17.16                 & 1.02                 & \multicolumn{1}{l}{17.26}
& \multicolumn{1}{l}{1.01} & \multicolumn{1}{l}{16.10}  & \multicolumn{1}{l}{1.02} & \multicolumn{1}{l}{19.39}    \\
\begin{tabular}[c]{@{}l@{}}Donor-recipient gender match\\  (baseline: F-F)\end{tabular} & \multicolumn{1}{r}{} & \multicolumn{1}{r}{} & \multicolumn{1}{r}{} & \multicolumn{1}{r}{}       & \multicolumn{1}{r}{}     & \multicolumn{1}{r}{}       & \multicolumn{1}{r}{}     & \multicolumn{1}{r}{}         \\
(F-M)                                                                                   & 1.08                 & 2.07  
& 1.08                 & 2.07                        & 1.07                     &  1.94                      & 1.14                     & 3.70                         \\
(M-M)                                                                                   & 0.89                 & -3.27                 & 0.89                 & -3.30                       & 0.88                     & -3.46                       & 0.94                     & -1.54                        \\
(M-F)                                                                                   & 1.01                 & 0.32                  & 1.01                 & 0.27                        & 1.01                     & 0.15                        & 1.08                     & 2.03                         \\
Recipient Obesity                                                                       & 1.16                 & 5.48                  & 1.16                 & 5.52
& 1.17                     & 5.62                        & 1.17                     & 5.62                         \\
Recipient Donor                                                                         & 1.05                 & 1.63                  & 1.05                 &1.65          
& 1.04                     & 1.46                        & 1.05                     & 1.65                         \\
\begin{tabular}[c]{@{}l@{}}Recipient race \\ (baseline : White)\end{tabular}            & \multicolumn{1}{r}{} & \multicolumn{1}{r}{} & \multicolumn{1}{r}{} &                             &                          &                             &                          &                              \\
(Black)                                                                                 & 1.52                 & 13.87                 & 1.52                 & 13.96                       & 1.50                     & 13.05                       & 1.53                     & 13.66                        \\
(Hispanic)                                                                              & 0.91                 & -2.22                 & 0.91                 & -2.06
& 0.98                     &  -0.34                      & 0.94                     &  -1.45 \\
(Other)                                                                                 & 0.80                 & -3.62                 & 0.80                 & -3.55                       & 0.84                     & -2.75                       & 0.80                     & -3.53                        \\
\begin{tabular}[c]{@{}l@{}}Donor race\\ (baseline : White)\end{tabular}                 & \multicolumn{1}{r}{} & \multicolumn{1}{r}{} & \multicolumn{1}{r}{} & \multicolumn{1}{r}{}       & \multicolumn{1}{r}{}     & \multicolumn{1}{r}{}       & \multicolumn{1}{l}{}     & \multicolumn{1}{l}{}         \\
(Black)                                                                                 & 1.30                 & 7.31                  & 1.30                 & 7.26                        & 1.28                     & 6.87                        & 1.32                     & 7.68                         \\
(Hispanic)                                                                              & 1.04                 & 0.92                  & 1.05                 & 1.04                        & 1.08                     & 1.69                        & 1.07                     & 1.50                         \\
(Other)                                                                                 & 1.07                 & 0.88                  & 1.08                 & 0.97                        & 1.14                     & 1.72                        & 1.12                     & 1.37                         \\
\begin{tabular}[c]{@{}l@{}}HLA mismatch  \\ (baseline: 0 MM 0 DR MM)\end{tabular}       & \multicolumn{1}{r}{} & \multicolumn{1}{l}{} & \multicolumn{1}{l}{} &                             &                          &                             &                          &                              \\
(1-2 MM 0 DR MM)                                                                        & 1.26                 & 2.99                  & 1.26                 & 3.07  
& 1.27                     & 3.11                        & 1.53                     & 5.88                         \\
(1-2 MM 1-2 DR MM)                                                                      & 1.51                 & 7.32                  & 1.51                 & 7.47
& 1.51                     & 7.33                        & 1.84                     & 12.25                        \\
(3-4 MM 0 DR MM)                                                                        & 1.34                 & 3.92                  & 1.35                 & 4.13  
& 1.40                     & 4.34                        & 1.60                     & 6.56                         \\
(3-4 MM 1-2 DR MM)                                                                      & 1.68                 & 9.91                  & 1.68                 & 10.09  
& 1.68                     & 9.89                        & 1.99                     & 14.55                        \\
Donor type (deceased)                                                                   & 1.78                 & 18.67                 & 1.78                 & 18.64
& 1.80                     & 18.69                       & 1.85                     & 19.58                        \\
Recipient diabetes                                                                     & 0.97                 & -0.99                 & 0.97                 &-0.98 
& 0.97                     & -1.08                       & 0.96                     & -1.30                        \\
Recipient hypertension                                                                 & 1.04                 & 1.29                  & 1.03                 & 0.99                        & 1.03                     & 0.96                        & 1.09                     & 2.63                         \\ 
\end{tabular}}}\label{tab:3}
\end{table}

Table \ref{tab:3} presents the estimated hazard ratios and their z-scores in the proportional hazards model across different methods. For the $\colsa$ method, we chose third-order Bernstein polynomials to approximate the log baseline hazard function. The estimates and standard errors obtained from \colsa\ and the oracle method are highly consistent across all covariates. However, the meta method generated incorrect inference results on the donor-recipient gender mismatch risk factor. Female donor and male recipient combination is significantly associated with higher risk for 5-year DCGF as shown from the oracle method, whereas the meta method underestimated both hazard ratio and its standard error, resulting in an insignificant estimate at the 95\% confidence level. Although Online method has consistent inference conclusions as the oracle method except for the Hispanic category of the recipient race variable, both point estimates and standard errors deviated noticeably more from the oracle estimates. Another clear disagreement between the results of the meta and Online methods and those of the oracle is the effect of being a Hispanic recipient. Compared to White recipients, Hispanic recipients have 9\% reduced risk for 5-year DCGF and the reduced risk is statistically significant for both oracle and \colsa\ methods which is consistent with the findings in literature~\citep{gordon2010disparities}. The same variable is not found to be statistically significant using meta or Online survival methods. Given that there is a known disparity in kidney transplant access among different racial demographics, the significant results should be interpreted with caution.
As evident in the results, little loss of statistical power occurred in \colsa, while overcoming data sharing barriers and enjoying the maximal protection of data privacy. 
It is also worth noting that our proposed \colsa\ approach is not affected by imbalanced distributions of covariates across study sites, a striking advantage over the classical meta-analysis method. The estimated baseline survival curves for the \colsa\ method, following the procedures described in Section \ref{proj2:sec:method:surv}, and for the oracle method using the Nelson–Aalen estimator, are shown in Figure \ref{proj3:figure:surv}.
\begin{figure}[h]
\includegraphics[width=8cm]{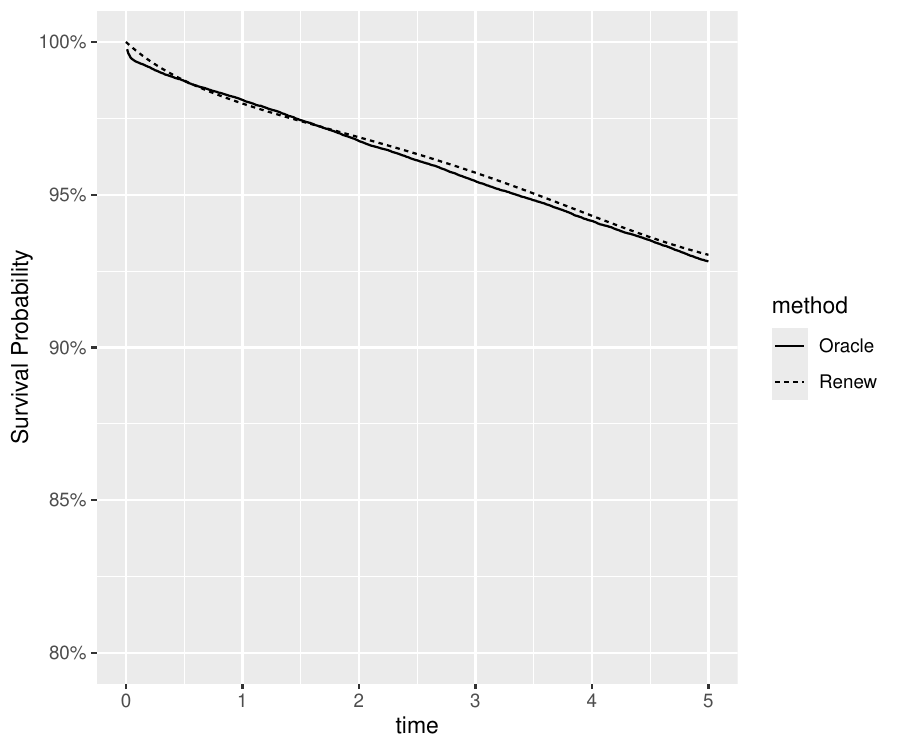}
\centering
    
	\caption{	 The estimated baseline survival curves obtained by the \colsa\ method and the oracle method.}

\end{figure}\label{proj3:figure:surv}

\section{Discussion} \label{sec: discussion}

This paper introduces a new methodology that overcomes data-sharing barriers and provides a privacy-enhanced efficient hazard ratio estimate in time-to-event data analyses. Unlike the approaches in the current literature on distributed Cox models, which face challenges of potential data leakage in risk-set construction, our method employs a sieve maximum likelihood approach with the baseline hazard estimated nonparametrically using Bernstein polynomials. We establish the desirable asymptotic properties of the proposed \colsa~method as well as investigate the finite-sample performance through simulation experiments. Our findings indicate that the proposed hazard ratio estimation is not sensitive to the choice of Berstein polynomial degree, so that our method enjoys flexibility in the baseline hazard approximation in the implementation. 
One minor limitation of our \colsa~method pertains to requiring good-quality data at the first site to yield consistent estimates. Similar to \cite{hu2024}, when each site has similar sample sizes and event rates, \colsa\ estimation results are expected to remain close over different orders of study sites pre-fixed for sequential updating.
Given that the first site's summary statistics will serve as the starting values for the subsequent update, we recommend starting the \colsa~procedure with the largest site or the site with the most events to avoid potential numerical issues (e.g. non-convergence) caused by poor initial values. 
{Another limitation of our \colsa\ method is that it is assumed to be operated under the condition that data collected across all sites are compatible with respect to the data generation distribution, meaning that hazard ratios and baseline hazards are homogeneous across sites. Some good diagnostic procedures to verify this homogeneity condition can be established by extending from the procedure introduced in \citep{hu2024} which checks the compatibility of parameters in the propensity score model. Sites with clear evidence of violating the homogeneity assumption should be discarded in the current form of our \colsa\ method. }Of note, our approach can be extended by using other types of polynomial splines with minimal effort. The R package for implementing \colsa\ is available at \emph{https://github.com/CollaborativeInference/COLSA}. 

The current \colsa\ methodology is developed for the low-dimensional Cox model. \cite{bayle2023communication} have introduced a distributed, high-dimensional, sparse Cox proportional hazards model, which, however, still suffers from vulnerability to information leakage in the risk-set construction. \colsa\ provides a possible solution to handle this issue, which is worth further exploration. To verify the proportional hazards assumption, \cite{xue2020online} proposed a divide-and-conquer test for testing the proportional hazards assumption with streams of survival data and future work can extend their approach for the \colsa\ method. In the case when there is evidence for potential violation of the proportional hazards assumption, \colsa\ method can be readily extended to accommodate the time-varying effects or consider a new development of accelerated failure modeling approach.

	\clearpage
	\spacingset{1.45}
	\bibliographystyle{agsm}
	\bibliography{SC}
	
\appendix
\begin{appendices}
\section{Additional regularity assumption}\label{proj2:appx1}
Denote the expectation of $f(y)$ under the distribution $P$ as $Pf = \int f(y)dP(y)$,  and denote the empirical expectation  of $f(y)$ 
 by $\mathbb{P}_n f = n^{-1}\sum_{i=1}^nf(y_i)$.
\begin{assumption}\label{sieve}

   The following regularity conditions are needed to establish the asymptotic properties for the oracle estimator. 
    \begin{enumerate}[label = (\alph*), ref = \ref{sieve}(\alph*)]

    \item \label{s1:a} 
    $\mathcal{B}$ is compact and contains the true parameter $\bbeta_0$ as its interior point.
    \item  $X$ belongs to a bounded subset of $\mathbb{R}^r$ and $E(XX^T)$ is nonsingular. 
     \item  There exists a truncation time $\tau < \infty$ such that, for some positive constant $\delta_0$, $P(Y > \tau|X) \geq \delta_0$ almost surely with respect to the probability measure of $X$.  
     \item \label{a1:continuity}
     {{The first-order derivative of the true log baseline hazard function $g_0$ is bounded and Lipschitz continuous.}}
    \item \label{a1:e} For some $\eta \in (0,1),$ $u^\T Var(X \mid Y,\Delta=1)u \geq \eta u^\T E(XX^\T \mid Y,\Delta=1)u$ almost surely for all $u\in R^r.$

          \item For every $\theta$ in $N_{\rho}(\btheta_0)$, $P\{l(\theta,\mathcal{O})-l(\theta_0,\mathcal{O})\}\leq -C \Vert\theta-\theta_0\Vert^2$ for some constant $C>0$.
    \end{enumerate}
\end{assumption}
\textit{Comment: }
Under \ref{a1:continuity}, within the space of all $p$-degree Bernstein polynomials $\mathcal{G}^p$, there exists a Bernstein polynomial $g_N$ such that {{$\Vert g_N - g_0 \Vert _{\infty}  = O(p^{-1/2})= O(N^{-\nu/2})$ (Theorem 1.6.2 in \cite{lorentz2012bernstein}; also see the proof of Lemma 2 in \cite{osman2012nonparametric})}}.

\vspace*{-10pt}

\subsection{Existing distributed proportional hazards model review}
We compare several relevant methods that primarily fit the Cox model through summary data sharing in Table~\ref{tab:methods}.  
\begin{table}[h]
\caption{
	A comparison of relevant methods. $d$ is the length of unique event times across all sites and $p$ is the number of covariates. $N$ is the total sample size.
	}\label{tab:methods}
 \resizebox{\columnwidth}{!}{
\begin{tabular}{@{}cllll@{}}
Authors                                                    & Method                                              & \multicolumn{1}{c}{\begin{tabular}[]{@{}c@{}}Event\\  times\end{tabular}}                               & \begin{tabular}[c]{@{}l@{}}Size of data\\ transferred \\ from each site\end{tabular} & Disadvantages                                                                                                                                                                                                                      \\  
\cite{lu2015webdisco}    & WebDISCO                                            & \begin{tabular}[c]{@{}l@{}}Public \\ information\end{tabular}                                            & \begin{tabular}[c]{@{}l@{}}$O(dp^2)$ for \\ each iteration\end{tabular}              & \begin{tabular}[c]{@{}l@{}}Inefficient because \\ it is iterative; \\ broadcast estimates\\ to local sites after\\  each iteration\\ till convergence\end{tabular}                                                                \\[1.5cm] 
\cite{duanlocaltoglobal} & ODAC                                                & \begin{tabular}[c]{@{}l@{}}Public\\ information\end{tabular}                                             & $O(dp^2)$                                                                            & \begin{tabular}[c]{@{}l@{}}Individual level data\\ can be exposed when \\ only a few individuals \\ are included in adjacent\\ time points; require\\ local HR estimate to \\ converge.\end{tabular}                              \\ [1.5cm] 
\cite{li2022distributed} & LLSTW                                               & \begin{tabular}[]{@{}l@{}}Public\\ information\\ or\\ Third Party\\ rank times\end{tabular}             & $O(dp^2)$                                                                            & \begin{tabular}[c]{@{}l@{}}Individual level data\\ can be exposed when \\ only a few individuals \\ are included in adjacent\\ time points; require\\ local HR estimate to \\ converge; a third party\\ is involved.\end{tabular} \\ [2cm]
\cite{lu2021multicenter}                              & \begin{tabular}[c]{@{}l@{}}LTZZL\end{tabular} & \begin{tabular}[c]{@{}l@{}}Public\\ information\end{tabular}                                             & Encrpyted Data                                                                       & \begin{tabular}[c]{@{}l@{}}Hard to implement for\\ practitioners;\\ require computational\\ intensive homomorphic \\ protocol.\end{tabular}                                                                                       \\[1.5cm] 
\cite{imakura2023dc}                                    & DC-cox                                              & \begin{tabular}[c]{@{}l@{}}Public \\ Information\end{tabular}                                            & $O(Np)$                                                                              & \begin{tabular}[c]{@{}l@{}}A shareable anchor \\ dataset is needed; a\\ third party is involved.\end{tabular}                                                                                                                     \\ [1.5cm] 
\cite{wu2021online}                                     & Online                                                & \begin{tabular}[c]{@{}l@{}}Private; can\\ deduce event \\ times when \\ partition is\\ fine\end{tabular} & $O(p^2)$                                                                             & \begin{tabular}[c]{@{}l@{}}Fixed partition does not\\ have optimal accuracy;\\ adaptive partition \\ requires large sample;\\ require local HR \\ estimate to converge.\end{tabular}                                              \\ [1.5cm] 
Our Method                                                 & \colsa                               & Private                                                                                                  & $O(p^2)$                                                                             &    \begin{tabular}[c]{@{}l@{}} Do not have the\\ disadvantages listed above.\end{tabular}                                                                                                                                                                                                                                 \\ 

\end{tabular}}
\end{table}

\end{appendices}

\end{document}